\documentclass[a4paper]{jpconf}
\usepackage{graphicx}
\begin{document}
\newcommand{\thp}{$\Theta^+$ }

\title{An Experimental Review of the $\Theta^+$ Pentaquark}

\author{Ken Hicks}

\address{Department of Physics and Astronomy, Ohio University, Athens OH 45701}

\ead{hicks@ohio.edu}

\begin{abstract}
Evidence for the \thp pentaquark is still sketchy at the present time. 
This state, if it exists, has a small width and consequently a small 
production cross section.  No single experiment has overwhelming 
evidence for the \thp and some non-observations of the \thp are 
difficult to understand unless this exotic baryon also has an 
exotic production mechanism.  However, new data from the LEPS and 
CLAS collaborations with higher statistics are on the way, which 
will likely resolve the question of whether the \thp exists.
\end{abstract}.

\section{Introduction}

The history of physics has many examples of experimental results 
that were controversial at first and later confirmed by better 
experiments \cite{gasiorowicz,pdg}.  Of course, there are also 
many examples of results that were later shown to be wrong with 
higher statistics \cite{barnes,close}.  The question of when a 
result is real (or when it is a mistake) is a tricky one, and 
the situation is exacerbated when people make strong statements 
based on scant data.  A better approach is to be patient and 
let science take its course.  If the results are real, then they 
will be borne out in time by better experiments.  In the 
meantime, it is best to be cautious when drawing conclusions. 

It is interesting to look at papers from the 1960's when new 
resonances were being discovered at a rather fast pace. For 
example, the first evidence for the $\omega$ meson \cite{maglic} 
showed a narrow peak on top of a broad phase space that had 
only $83 \pm 16$ counts (about 5-$\sigma$) on top of a background 
of about 98 counts (see Fig. 1).  In this measurement, the width of 
the peak is about 30 MeV, which is close to the experimental resolution 
of 24 MeV.  Within two years, there were many papers \cite{gelfand} 
confirming this result with higher statistical significance 
and better resolution.

The situation with the \thp (an exotic baryon with a 4-quark 
plus one anti-quark $udud\bar{s}$ valence structure) is not so 
unlike the experimental evidence of the $\omega$ meson, except 
that the statistics are a bit lower (see Fig. 2).  The first 
papers (reviewed in Ref. \cite{nakanohicks}) reported 
a narrow peak \cite{leps,diana,clas-d} 
with about 4-5 $\sigma$ statistical significance
on top of a broad background.  At the 
time of writing, it has been a year plus a few months since the 
first publication of the LEPS collaboration, and there have been 
10 experimental papers confirming the \thp (see Table 1) with 
statistical significance ranging from 3 to 7 $\sigma$.  What is 
different from the case of the $\omega$ meson is that the \thp 
baryon is not seen in some experiments (high-energy, where production 
by fragmentation is dominant, nor is it seen in $e^+e^-$ collisions 
at B-factory energies where other baryon-antibaryon 
resonances are seen).  These non-observations are worrisome, and 
suggest that if the \thp exists, its production mechanism is 
somehow suppressed (in an unexpected or ``exotic" way).  This 
leads to theoretical speculations and creates doubt that the 
\thp exists.

\begin{figure}[ht]
\begin{minipage}{18pc}
\includegraphics[width=18pc]{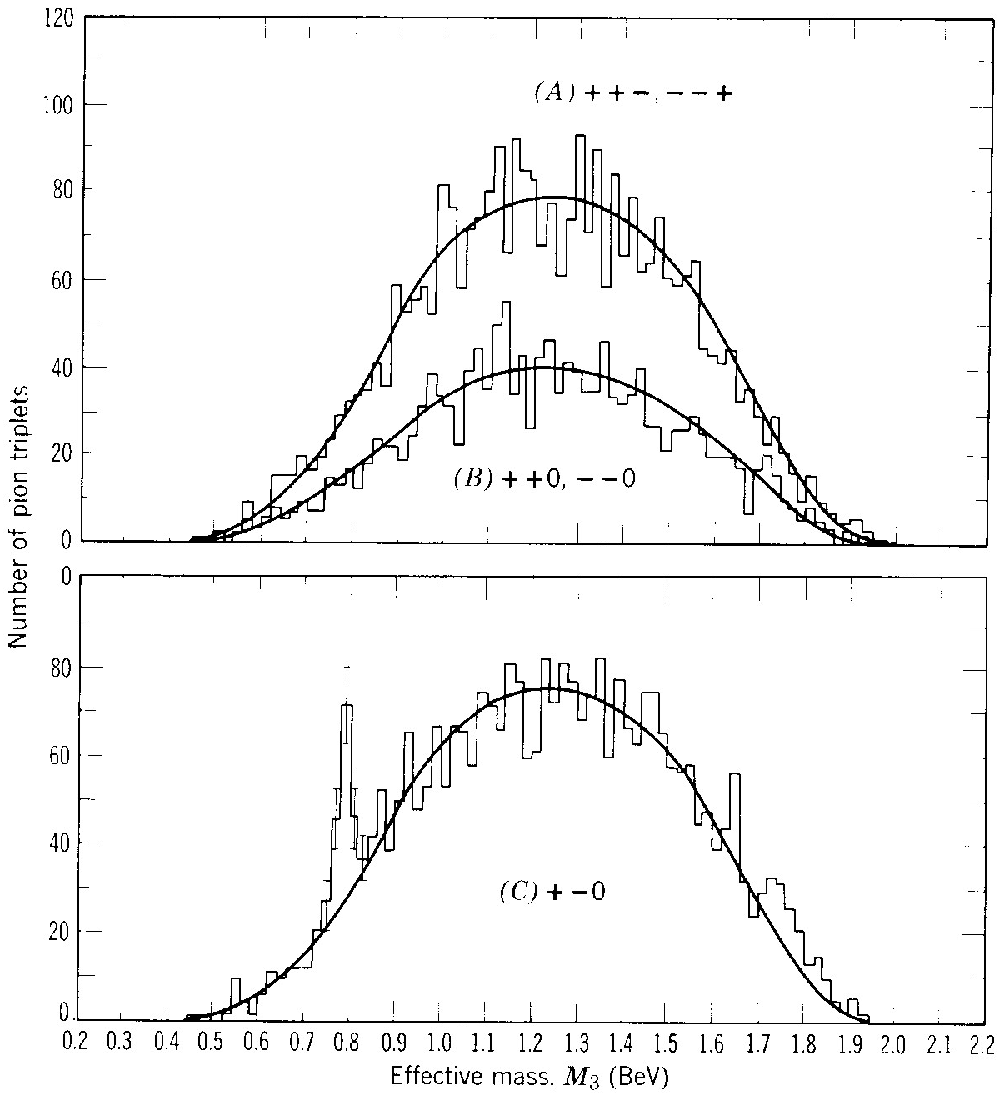}
\caption{\label{fig:omega} First evidence for the discovery of the $\omega$ 
	vector meson \cite{maglic}. The plots show the invariant mass 
	distribution for 3 pions with two having the same charge (top) or 
	one of each type ($+$, $-$ or 0 charge), showing that the $\omega$ 
	is an isoscalar particle.}
\end{minipage}\hspace{2pc}%
\begin{minipage}{18pc}
\includegraphics[width=18pc]{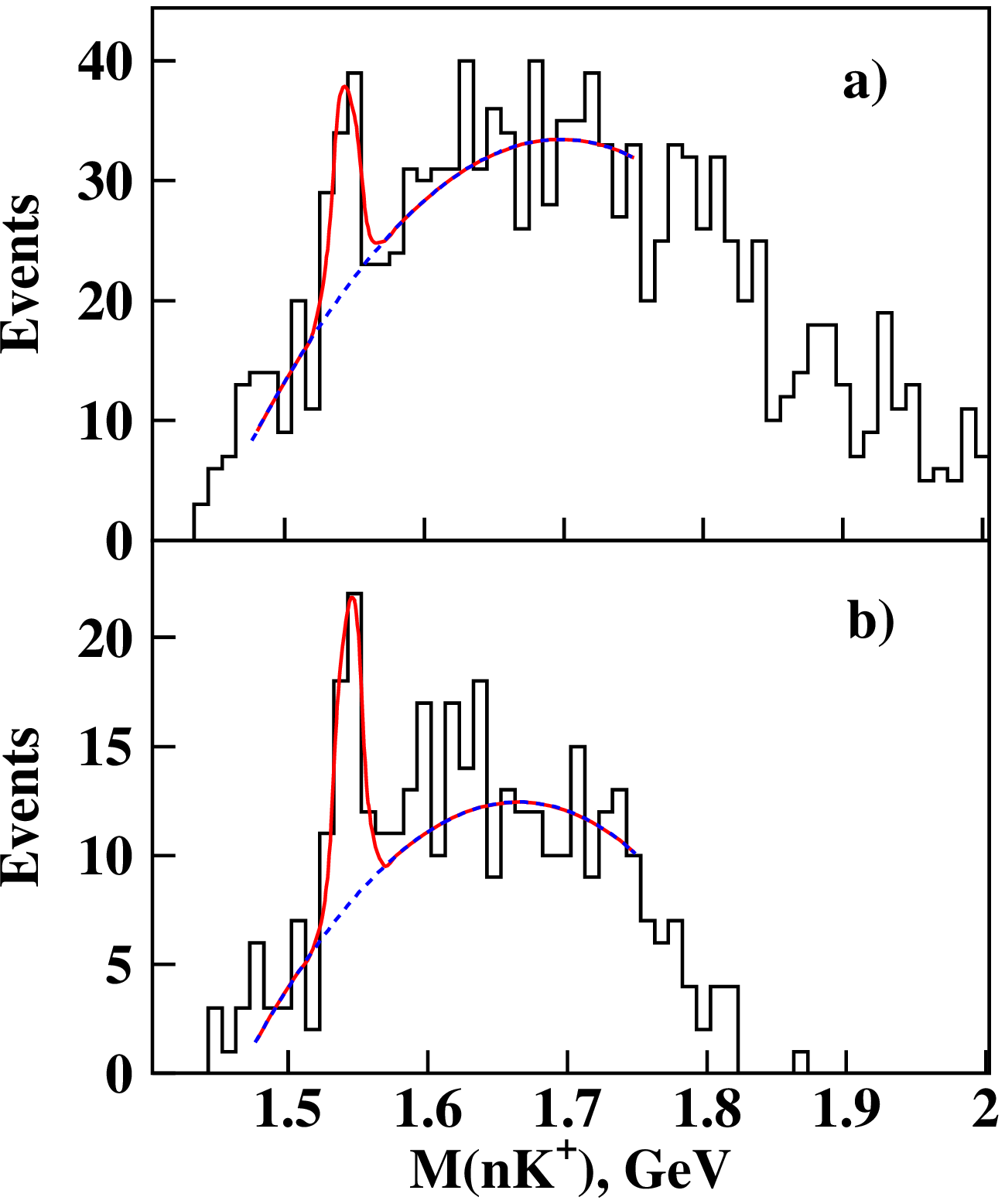}
\caption{\label{fig:clas-d}Evidence for the \thp pentaquark from CLAS data 
	on deuterium \cite{clas-d} with loose (top) and tight (bottom) 
	analysis cuts.}
\end{minipage} 
\end{figure}

In fact, if the \thp does not exists, it will be an interesting 
case for the history of physics.  Did the experimental reports 
underestimate the statistical significance?  Is there some 
kind of kinematic reflection, combined with statistical 
fluctuations, that could create this effect?  If so, could such 
an effect explain all of the \thp experimental results?
In the next two sections, some of these questions will be examined.  
In the last section, an outlook for future experiments 
with better statistics will be presented.

\section{Experimental Evidence}

Experiments with positive evidence for the \thp are shown in 
Table 1.  The mass spectra for these experiments have been shown 
at many conferences \cite{nakanohicks}, and so are not repeated here.  
There are several points to make regarding these positive results:
\begin{itemize}
\item There are many different reaction mechanisms that are 
listed in the table.  While it is possible to construe a kinematic 
reflection \cite{dzierba} or a cusp effect that might 
affect one reaction channel, it is very difficult to find a 
mechanism that would create a false peak in all cases.
\item Each of these experiments has some weakness \cite{hadron03}.
For example, some experiment require harsh cuts that, combined 
with detector acceptance, could possibly create unexpected 
structures in the mass spectra.  In other cases, the shape of 
the background under the peak is not well known, which could 
affect the statistical significance.  
\item The statistical significance in the table is calculated 
as a fluctuation of the background.  This is {\it not} the same 
procedure as used to calculate the area under a peak.  Since 
the shape of the background is not known perfectly, the 
$\sigma$'s shown in the last column are likely an overestimate 
of the statistical significance \cite{hermes}.
\item The masses are a bit inconsistent.  Taken as a whole, 
almost all measurements are within 1-2 standard deviations 
of the average value, 1535 MeV.  However, some measurements 
are clearly inconsistent with others, such as the DIANA and 
ZEUS results.  Either the experimentalists have underestimated 
their systematic errors, or there is a serious problem to be 
faced (suggesting that the \thp might not exist, or some 
other explanation such as a weak, yet-unknown $\Sigma^+$ 
resonance near 1525 MeV).
\item No single experiment makes a really convincing case that 
the \thp exists.  What is needed is a really good experiment 
with high statistics that shows a strong peak over a smooth 
background.  Until this happens, we must take a cautious view 
on the existence of the $\Theta^+$.
\end{itemize}

\begin{center}
\begin{table}[ht]
\caption{Published experiments with evidence for the \thp baryon.}
\centering
\begin{tabular}{|c|l|c|c|c|c|}
\br
Reference& Group& Reaction 	& Mass 		& Width &$\sigma$'s* \\
	&	&		& (MeV)		&(MeV)	&   \\
\hline
\cite{leps} &
LEPS	& $\gamma C \to K^+ K^- X$	& $1540\pm 10$	& $<25$	& 4.6 \\
\cite{diana} &
DIANA	& $K^+ Xe \to K^0 p X$		& $1539\pm 2$	& $<9$	& 4.4 \\
\cite{clas-d} &
CLAS	& $\gamma d \to K^+ K^- p (n)$	& $1542\pm 5$	& $<21$	&$5.2\pm 0.6^\dagger$\\
\cite{saphir} &
SAPHIR	& $\gamma d \to K^+ K^0 (n)$	& $1540\pm 6$	& $<25$	& 4.8 \\
\cite{itep} &
ITEP	& $\nu A \to K^0 p X$		& $1533\pm 5$	& $<20$	& 6.7 \\
\cite{clas-p} &
CLAS	& $\gamma p\to\pi^+ K^+K^-(n)$	& $1555\pm 10$	& $<26$	& 7.8 \\
\cite{hermes} &
HERMES	& $e^+ d \to K^0 p X$		& $1526\pm 3$	& $13\pm 9$&$\sim 5$\\
\cite{zeus} &
ZEUS	& $e^+ p \to e^+ K^0 p X$	& $1522\pm 3$	& $8\pm 4$& $\sim 5$\\
\cite{cosy} &
COSY-TOF& $p p \to K^0 p \Sigma^+$	& $1530\pm 5$	& $<18$	& 4-6  \\
\cite{svd} &
SVD	& $p A \to K^0 p X$		& $1526\pm 5$	& $<24$	& 5.6  \\
\br
\end{tabular}
\\
$^*$ Gaussian fluctuation of the background, as $N_{peak}/\sqrt{N_{BG}}$.
This ``naive" significance may underestimate the real probability of 
a fluctuation by about 1-2 $\sigma$.\\
$^\dagger$ Further analysis of the CLAS deuterium data suggest that the 
significance of the observed peak may not be as large as indicated.
\end{table}
\end{center}

What is not shown in the table are the beam energies used.  Many of 
these experiments are done near (within a few GeV) to the threshold for 
\thp production.  This provides an advantage, because it limits the 
number of possible reactions that can contribute to the background. 
For example, the LEPS data has a maximum beam energy of 2.4 GeV, 
which is too low for production of higher-mass mesons (such as the 
$a_2$(1320) followed by decay to a $K^+K^-$ pair with enough energy 
to be detected).  At higher energies, many more particles with high 
masses are produced and follow decay paths that are not well known. 
Determining the background shape is easier for near-threshold 
experiments, such as COSY-TOF, where a limited number of calculable 
cross sections contribute to the background shape under the \thp peak.

In any measurement, there is a chance that the production of other 
particles can ``reflect" into the mass spectrum of interest.  A 
possible mechanism has been described by Dzierba \etal \cite{dzierba}. 
However, such reflections are more likely to create broad peaks (widths 
of 50-100 MeV) rather than narrow peaks (widths of 20 MeV or so). 
Still, broad peaks coupled with low statistics can cause fluctuations 
that might result in narrow peaks.  Simulations of these processes, 
coupled with the detector acceptance and the analysis cuts used are 
an essential step in a good experiment.  For example, the CLAS 
experiment \cite{clas-p}, as well as others shown in the table, 
did these necessary simulation studies.  The specific model used by 
Dzierba \etal has been refuted \cite{hicks}, and is no 
longer of concern for the CLAS analysis, but in general one must 
be careful to consider the effect of kinematic reflections.

Assuming for the moment that the \thp is real, then a theoretical 
model is needed to explain its structure.  The \thp was predicted 
by Diakonov, Petrov and Polyakov (DPP) \cite{dpp} at a mass of 
about 1530 MeV with a width of $<15$ MeV.  One prediction of the DPP 
model is that the \thp is part of a group structure (a $\overline{10}$) 
and other exotic baryons, such as the $\Xi^{--}$ should also 
exist.  One experiment \cite{na49} has claimed to see the $\Xi^{--}$ 
at a mass of about 1860 MeV, but there has been no confirmation 
of this result by other experiments.  If the other members of the 
$\overline{10}$ group are not found, then this brings into doubt 
the interpretation of the \thp within this model.  Again, we must 
be cautious about depending too heavily on one model, and ask 
whether the \thp could have some other interpretation, such as 
a bound $\pi KN$ state \cite{bicudo,kishimoto} or some other theory. 
The important thing is to continue experimental searches for the 
\thp and for other members of the $\overline{10}$ group.

\section{Published Non-observations}

Experiments having a null result in searches for the \thp are listed 
in Table 2.  Again, these mass plots have been shown before, 
and will not be repeated here.  In the positive results of the 
previous section, some criticisms were listed.  In fairness, we 
should also be critical of the non-observations:
\begin{itemize}
\item These experiments break down into two categories: $e^+e^-$
production, where there are no quarks in the initial state, 
and high-energy proton collisions, where the multiplicity of 
particles in the final state is typically large.

\item The production mechanism of the \thp (if it exists) is not 
known. In the case of $e^+e^-$ annihilation, baryon-antibaryon 
production has a lower probability than meson pair production 
(see below).  No reasonable theoretical prediction exists for 
the probability of $\Theta \bar{\Theta}$ production.  Comparisons 
with, say, $\Lambda (1520)$ production are of limited use 
without theory to guide us.

\item Similarly, the production mechanism of the \thp from 
high-energy proton scattering, which proceeds mainly through 
fragmentation of the projectile or the target, is unclear. 
One theoretical estimate \cite{titov} shows that the 
\thp production in the fragmentation model of quark counting 
rules calculates that the ratio of \thp production is 
suppressed by more than $10^3$ as compared with $\Lambda (1520)$ 
production. (The limits shown in the table are typically 
a few percent of $\Lambda^*$ production rates.)

\item When the particle multiplicities are high, the method 
to determine which particle is produced at the same vertex 
as the detected $K_s^0$ becomes difficult.  For example, if 
there are 5 protons in the same event as the $K_s^0$, then 
all five combinations must be used for the invariant mass 
of the $pK^0$ spectrum, unless some of these protons can 
be identified with another particle in the event. This 
combinatorial background can be significant, and one should 
look carefully at how these backgrounds are handled in 
the non-observation experiments. 
\end{itemize}

\begin{center}
\begin{table}[ht]
\caption{Published experiments with non-observation of the \thp baryon.}
\centering
\begin{tabular}{|c|l|c|c|c|}
\br
Reference& Group& Reaction 	& Limit 	& Sensitivity? \\
	&	&		& 		&	   \\
\hline
\cite{bes} &
BES	& $e^+e^- \to J/\Psi \to \bar{\Theta}\Theta$	& $<1.1\times 10^{-5}$ B.R.	& No$^*$  \\
\cite{belle} &
Belle	& $e^+e^- \to \bar{B}^0 B^0 \to \bar{p}pK^0$	& $<2.3\times 10^{-7}$ B.R.	& $\Theta^{++}$  \\
\cite{babar} &
BaBar	& $e^+e^- \to \Upsilon (4S) \to pK^0 X$		& $<1.0\times 10^{-4}$ B.R.	& ??  \\
\cite{hera-b} &
HERA-B	& $p A \to K^0 p X$		& $<0.02 \times \Lambda^*$ 	& No?  \\
\cite{cdf} &
CDF	& $p \bar{p} \to K^0 p X$	& $<0.03 \times \Lambda^*$ 	& No?  \\
\cite{phenix} &
PHENIX	& $Au+Au \to K^- \bar{n} X$	& (not given) 	& ??  \\
\br
\end{tabular}
\\
$^*$ see Ref. \cite{azimov} for calculations.
\end{table}
\end{center}

The detection of the complete final state in exclusive reactions 
holds an advantage over the inclusive production presented in 
the non-observation publications.  This is especially true for 
experiments at near-threshold production, where the number of 
particles in the final state are limited.  In the high-energy 
experiments, the production of hadrons is thought to go via a 
complicated ``hadronization" process of string-breaking and 
statistical energy sharing \cite{lund}.  When a high-mass 
baryon is produced, it will go through subsequent decays that 
may preferentially populate one state, such as the $\Lambda (1520)$ 
as opposed to, say, the $\Sigma^0 (1660)$. Without theory to 
guide us, systematics of various baryon final states need to be 
studied with the goal to estimate the uncertainties in baryon 
production mechanisms.

\begin{figure}[ht]
\includegraphics[scale=0.8]{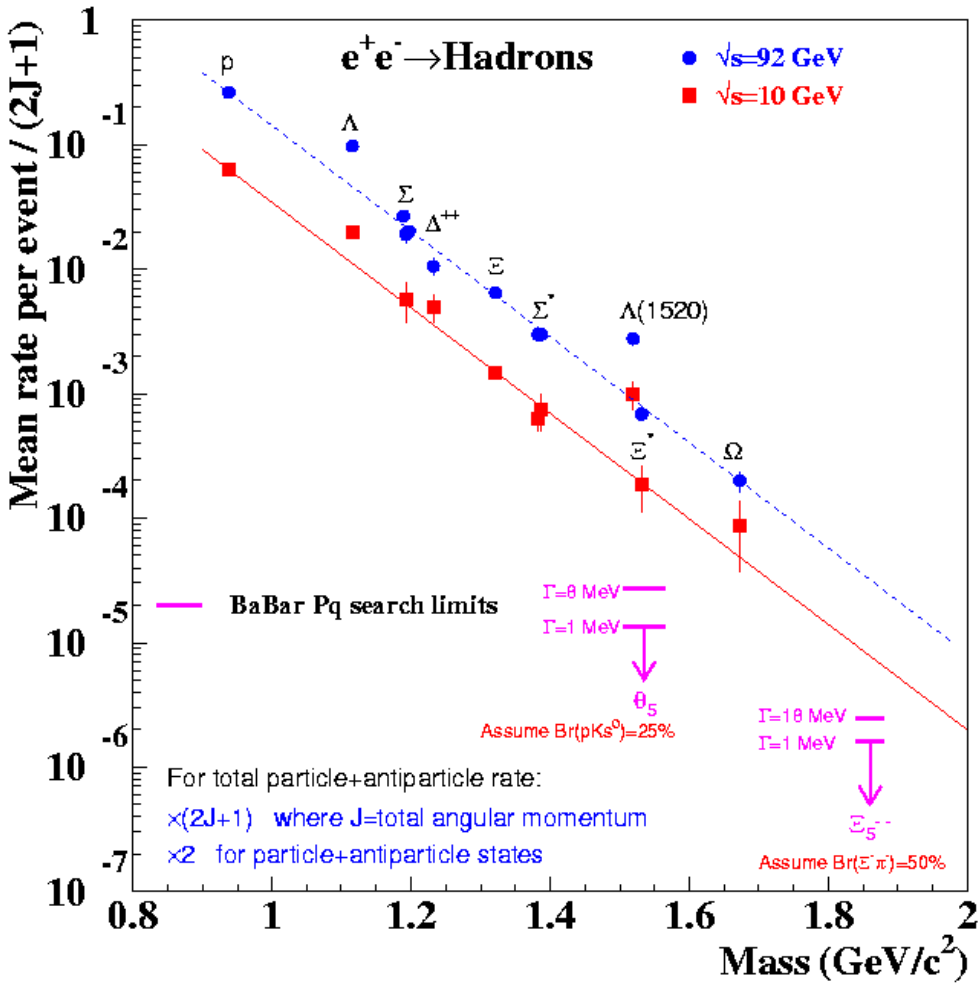}
\caption{\label{fig:babar} Phenomenology of the rate hadron production 
from $e^+e-$ collisions the BaBar experiment \cite{babar}.}
\end{figure}

The BaBar experiment \cite{babar}  has investigated some of 
the systematics of meson and baryon production from $e^+e^-$ 
collisions, as shown in Fig. \ref{fig:babar}. There are several 
interesting points regarding this plot. First, there is a different 
slope (by a factor of two) for mesons and baryons \cite{halyo}.  
Of course, mesons 
have a quark-antiquark pair and baryons have 3-quarks, so it 
is easy to believe that their respective production rates 
should be different.  An extension of this line of reasoning 
suggests that {\it pentaquark particles should also have yet a 
different slope}. Again, theoretical models are needed to 
calculate the slope for pentaquark production. Until this 
is done, we do not know if the \thp should have been seen 
in these experiments. Second, it is interesting to see that 
some baryons fall significantly above or below the average 
of the lines in Fig. \ref{fig:babar}.  
For example, the $\Lambda (1520)$ production 
rate is about 3-4 times higher than one would expect based 
on the systematics.  Similarly, the $\Lambda$ ground state 
rate is about 2 times higher.  So far, no theory can explain this. 
Clearly, there is a lot of theoretical work necessary before 
the expected rate of \thp production in $e^+e^-$ collisions 
can be calculated.

An essential ingredient to the argument that the non-observations 
imply a non-existence of the \thp is that the production 
probability of pentaquark particles is similar to the production 
probability of 3-quark baryon resonances. As just discussed, there 
is good reason to doubt this assumption. Hence, we are left with 
a situation where the non-observations of the \thp are not 
convincing negative evidence, and the low-statistics experiments 
of the previous section are not convincing positive evidence. 
Obviously, the next step is to get higher statistics in an experiment 
where positive results were already seen. 

\section{Experimental Outlook}

Several collaborations are pursuing higher-statistics experiments 
with the goal to determine the existence (or not) of the $\Theta^+$. 
At HERMES, by the end of 2004 they will collect twice the data as 
used in their publication \cite{hermes}.  At COSY-TOF, an upgrade 
will allow better vertex resolution and overall they expect about 
5 times the data under similar conditions to Ref. \cite{cosy}.
At KEK, a new high-resolution experiment to measure 
$K^+ p \to \Theta^+ \pi^+$ 
has been approved \cite{imai} and will run in early 2005.
At LEPS, data from a deuterium target (with more statistics by a 
factor of a few) was taken in 2003 and has already been shown at 
various conferences and is being readied for publication.  At CLAS, 
new high-statistics data \cite{hicstep} similar to conditions of 
Ref. \cite{clas-d} (and also data on a proton target \cite{infn}) 
were completed in mid-2004, and are currently under analysis.
An outlook of the latter two experiments will be given below.

At SPring-8, 
data on a 15 cm liquid deuterium target were taken during a few 
months in 2003 at a photon intensity of about $10^6$/sec.  The 
reaction of interest is $\gamma d \to K^+ K^- X$ and under the 
assumption of a spectator proton, the missing mass of the $K^+$ 
gives $Y^*$ resonances with strangeness $S=-1$, such as 
$\Lambda (1520)$, and the missing mass of the $K^-$ gives possible 
\thp resonances with strangeness $S=+1$.  The LEPS detector 
covers only the forward angles \cite{leps} and is symmetric in 
acceptance for $K^+$ and $K^-$ particles.  Because of Fermi 
momentum, a kinematic correction is necessary \cite{nakano} 
which is applied in the same way for both $Y^*$ and \thp mass 
spectra.  These data have been shown at other conferences 
\cite{leps-d} and preliminary analysis indicates a peak with 
more statistical significance 
than the previous publication \cite{leps} but refinements of the 
analysis are still in progress.  Final results are expected to 
be submitted for publication in early 2005.

Jefferson Lab experiment 03-113 \cite{hicstep} was run during March 
to May of 2004 at an electron beam energy of 3.776 GeV.  Two separate 
magnetic field settings of the CLAS spectrometer \cite{mecking} 
were taken, one at 80\% of the maximum (same as Ref. \cite{clas-d}) 
and one at 60\%.  The latter field setting was done to reduce 
the loss of forward-angle $K^-$ particles, which are bent into 
the ``hole" of the CLAS acceptance (at the exit of the beam pipe). 
The two magnetic field settings also provide a consistency check 
because particles of the same momentum will traverse different 
paths for different B-field settings. The integrated luminousity 
at each B-field setting was approximately 10 times greater than 
that of the earlier publication \cite{clas-d}, although about 
half of the photon flux is at higher energy than used earlier.

\begin{figure}[ht]
\includegraphics[scale=0.75]{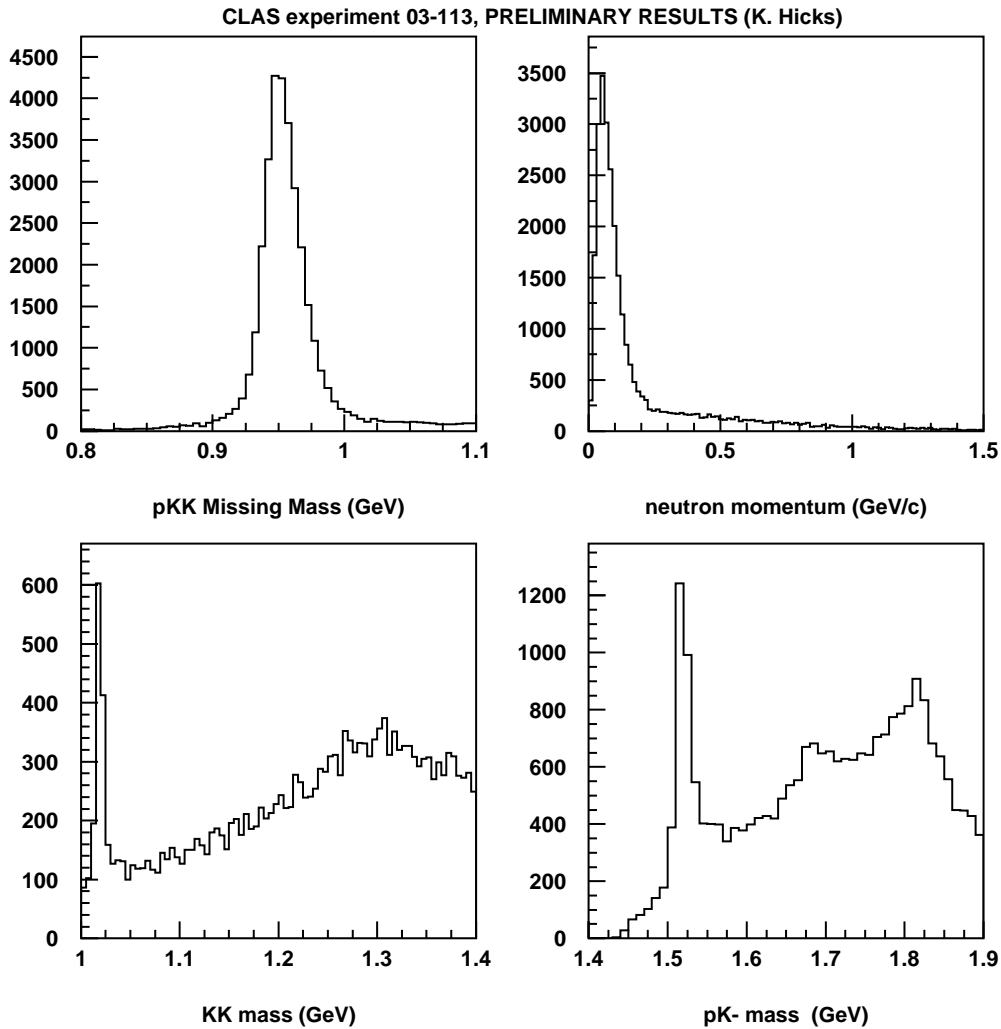}
\caption{\label{fig:g10} Mass spectra from about half of the 
high B-field data from CLAS for the high-statistics 
experiment 03-113 on a deuterium target.}
\end{figure}

Mass spectra from about 50\% of the high-field data from the 
CLAS experiment are shown in Fig. \ref{fig:g10} for the reaction 
$\gamma d \to K^+ K^- p (n)$.  The neutron is deduced by the 
missing mass of the $K^+K^- p$ final state, as shown in the 
top left plot.  These spectra are {\it still preliminary} and 
are integrated over all photon energies and all particle angles 
(see Ref. \cite{rossi}).
The resolution in these plots will be improved, and refinements 
of calibrations and rejection of background (such as mis-identified 
particles) are still in progress.  The distribution of neutron 
momenta are shown by the top right plot, for a cut on the 
$MM(K^+K^-p)$ from 0.90 to 0.98 GeV, and shows a peak at the 
Fermi momentum corresponding to reactions where the neutron is 
a spectator.  In the final analysis for \thp production, 
the neutron from \thp decay is expected to have momentum above 
0.2 GeV/c and this event selection will reduce the background 
from neutron spectator reactions.  Invariant mass spectra for 
the detected $K^+K^-$ and $pK^-$ pairs are shown in the lower 
two plots.  The $\phi$(1020) peak is clearly seen, along with 
a broad background that rises and falls in part due to the 
detector acceptance (these raw spectra are uncorrected for the 
acceptance).  In the $pK^-$ mass, the $\Lambda$(1520) shows a 
strong peak, along with broader peaks at about 1.68 and 1.81 GeV 
due to higher-mass $\Lambda^*$ states.  We note that these 
higher-mass states were not seen in our previous analysis 
\cite{clas-d} due to a lower photon beam energy. Also, these 
states are not seen in the non-observation experiments given 
in Table 2, which suggests that there are important differences 
in reaction mechanisms between these experiments and those 
done at CLAS.

\section{Summary}

There will continue to be critics of the evidence for the 
\thp pentaquark, as long as the real statistical significance
(as opposed to the naive one) is relatively low, or if 
severe angle cuts have been applied to the data.  The situation 
is complicated by the fact that high-energy experiments have 
not observed the \thp. However it is not clear if the \thp can 
be seen in fragmentation, where constituent counting rules 
suggest a substantial suppression of its production rate 
\cite{titov}. Because of the uncertainty in the production 
mechanism, the non-observation experiments do not rule out 
the existence of the \thp any more than they rule out the 
existence of the $\Sigma$(1660) which is also not seen 
in the $pK^0$ spectra of these experiments.

Kinematic reflections \cite{dzierba} have been refuted in 
the case of the CLAS data \cite{hicks} and seem unlikely 
to explain the \thp peaks in the 10 experiments with 
positive evidence, which use many different reactions. 
However, sources of background in all experiments (both 
those with observations of the \thp and those without) 
should continue to be investigated.  Some experiments 
have already done careful simulations, but more theoretical 
input for these simulations is desired.

High-statistics experiments using medium-energy probes at 
near-threshold production are the key to solving this dilemma. 
Several experiments have been done and are under analysis. 
It is likely that the question of whether the \thp exists or 
not will be answered by mid-2005.

\section*{Acknowledgments}

I am thankful for discusssions with many colleagues, both 
experimental and theoretical. Special thanks go to my colleagues 
of the LEPS and CLAS collaborations, and in particular to Takashi 
Nakano, Tsutomu Mibe, Stepan Stepanyan, Volker Burkert and Daniel 
Carman, for their many contributions to these experiments.

\end{document}